\documentstyle[epsf,sprocl]{article}
\def\Journal#1#2#3#4{{#1} {\bf #2}, #3 (#4)}

\def\npa{{\em Nucl. Phys.} A}

\def\pl{{\em Phys. Lett.}  B}
\def\prl{\em Phys. Rev. Lett.}
\def\prd{{\em Phys. Rev.} D}
\def\prc{{\em Phys. Rev.} C}
\def\pr{\em Phys. Rev.} 
\def\ptp{\em Prog. Theor. Phys.}

\def\jetp{\em Sov. Phys. JETP~}
\def\epl{\em Europhys. Lett.}
\def\jetpl{\em JETP Lett.}
\def\nima{{\em Nucl. Instrum. Methods} A}
\def\zpa{{\em Z. Phys.} A}
\def\sjnp{\em Sov. J. Nucl. Phys.}
\def\anp{\em Adv. Nucl. Phys.}

\def\bea {\begin{eqnarray}}
\def\eea {\end{eqnarray}}
\def\be {\begin{equation}}
\def\ee {\end{equation}}
\def\ben{\begin{enumerate}}
\def\een{\end{enumerate}}
\def\bi{\begin{itemize}}
\def\ei{\end{itemize}}
\def\ie{{\it i.e.}}

\def\etal{{\it et al.}}

\def\F{{\cal F}}

\def\GA{G_{\mbox{\tiny A}}}
\def\GV{G_{\mbox{\tiny V}}}
\def\GF{G_{\mbox{\tiny F}}}
\def\DRV{\Delta_{\mbox{\tiny R}}^{\mbox{\tiny V}}}
\def\DRA{\Delta_{\mbox{\tiny R}}^{\mbox{\tiny A}}}

\def\mA{m_{\mbox{\tiny A}}}

\def\mZ{m_{\mbox{\tiny Z}}}

\def\mids{\! \mid \! }

\newcommand{\sfrac}[2]{\mbox{\small{$\frac{#1}{#2}$}}}

\begin{document}

\title{THE EVALUATION OF $V_{ud}$, EXPERIMENT AND THEORY}
\author{I. S. TOWNER}
\address{Physics Department, Queen's University, 
Kingston, \\ Ontario K7L 3N6, Canada \\
E-mail: towner@sno.phy.queensu.ca}
\author{J. C. HARDY}
\address{Cyclotron Institute, Texas A \& M University, \\ 
College Station, TX 77843 \\
E-mail: hardy@comp.tamu.edu}
\maketitle
\abstracts{
The value of the $V_{ud}$ matrix element of
the Cabibbo-Kobayashi-Maskawa (CKM) matrix can be
derived from
nuclear superallowed beta decays, neutron decay, and pion
beta decay.  We survey current world data for all three. 
Today, the most precise value of $V_{ud}$ comes from the
nuclear decays; however, the precision is limited not by
experimental
error but by the estimated uncertainty in theoretical
corrections.  Experimental uncertainty does limit the
neutron-decay result, which, though statistically consistent
with the nuclear result, is approximately a factor
of three poorer in precision.
The value obtained for $V_{ud}$ leads to a result that
differs at the 98\% confidence level from the
unitarity condition for the  
CKM matrix.  We examine the
reliability of the small calculated corrections that
have been applied to the data, and assess the
likelihood of even higher quality nuclear data
becoming available to confirm or deny the
discrepancy.  Some of the required experiments
depend upon the availability of intense radioactive
beams.  Others are possible today.
}

\section{Introduction} \label{intro}

The Cabibbo-Kobayashi-Maskawa matrix\cite{Ca63,KM73} 
relates the quark eigenstates of
the weak interaction with the quark mass eigenstates (unprimed)

\be
\left ( \begin{array}{c} d^{\prime} \\ s^{\prime} \\ b^{\prime}
\end{array} \right ) =
\left ( \begin{array}{ccc} V_{ud} & V_{us} & V_{ub} \\
                           V_{cd} & V_{cs} & V_{cb} \\
                           V_{td} & V_{ts} & V_{tb} 
\end{array} \right ) 
\left ( \begin{array}{c} d \\ s \\ b
\end{array} \right )  
\label{CKM}
\ee

\noindent and, as such, the matrix is unitary.  Thus there are many 
relationships among the nine elements of the matrix that can be tested 
by experiment.  The leading element, $V_{ud}$, only depends on quarks 
in the first generation and so is the element that can be determined
most precisely.  Here we will discuss the current status of $V_{ud}$ 
and the unitarity test as it relates to the elements in the first row:

\be
V_{ud}^2 + V_{us}^2 + V_{ub}^2 = 1 .
\label{Unitarity}
\ee

In examining the unitarity test, we will adopt the Particle
Data Group (PDG02) \cite{PDG02}
recommendations for $V_{us}$ and $V_{ub}$.  We note
that for $V_{us}$ PDG02 recommends  
only the value determined from $K_{e3}$ decay, $ \, \mids 
V_{us} \mids \, = 0.2196 \pm
0.0023$, arguing that the value obtained from hyperon
decays suffers from
theoretical uncertainties due to first-order SU(3)
symmetry-breaking effects in the axial-vector couplings.
It now appears, though, from preliminary results \cite{Sh02}
that the experimental result for the e3
branch of the $K^+$ decay may also be in doubt.  For
the time being, we continue to use the value of
$V_{us}$ from PDG02 but acknowledge that this result
may have to change in future.
As to $V_{ub}$, its value is small,
$ \mids V_{ub} \mids \, = 0.0036 \pm 0.0010$, and
consequently it has a
negligible impact on the unitarity test, Eq.\
(\ref{Unitarity}).

\section{The value of $V_{ud}$} \label{vVud}

The value of $V_{ud}$ can be determined from three
distinct sources:
nuclear superallowed Fermi beta decays, the decay of the
free neutron,
and pion beta decay.  We discuss each in turn.

\subsection{Nuclear superallowed Fermi beta decays}
\label{nsFd}

Nuclei have the singular advantage that transitions with
specific characteristics can be selected and then isolated
for study. One example is the superallowed $0^{+}
\rightarrow 0^{+}$ beta transitions, which depend
uniquely on the vector part of the weak interaction. 
Furthermore,
in the allowed approximation, the nuclear matrix element
for these transitions is given
by the expectation value of the isospin ladder operator,
which is
independent of any details of nuclear structure and is
given simply
as an SU(2) Clebsch-Gordan coefficient.  Thus, the
experimentally
determined $ft$-values are expected to be very nearly the
same for all
$0^{+} \rightarrow 0^{+}$ transitions between states of a
particular     
isospin, regardless of the nuclei involved. Naturally, there
are corrections to this simple picture coming from
electromagnetic effects, but these corrections are small --
of order 1\% -- and calculable.  Thus, if we write
$\delta_R$ as the nucleus-dependent part of the radiative
correction, $\DRV $ as the nucleus-independent part of
the radiative
correction, and $\delta_C$ as the isospin symmetry-breaking
correction, then the experimental $ft$-value can be
expressed as follows:

\be
ft (1 + \delta_R )(1 - \delta_C ) \equiv \F t =  
\frac{K}{2 \GV^2 (1 + \DRV )} = ~{\rm constant} ,
\label{Ftconst}
\ee

\noindent where $K =
2 \pi^3 \ln 2 \hbar (\hbar c)^6 / (m_e c^2)^5$, which has
the value
$K/(\hbar c)^6 = ( 8120.271 \pm 0.012) \times 10^{-10}$
GeV$^{-4}$s; and $\GV$ is the weak vector coupling
constant: $\GV = \GF V_{ud}$, with $\GF$ being the weak
coupling constant from purely leptonic muon decay.
Thus, to extract $V_{ud}$ from experimental data, the
procedure is to determine the $\F t$-values for a
variety of different nuclei having the same isospin, and then
to test if they are self-consistent.  If they are, their average
is used to determine a value for $\GV$ and, from it,
$V_{ud}$.

{\footnotesize
\begin{table} [t]
\begin{center}
\caption{Experimental results ($Q_{EC}$, $t_{1/2}$ and
branching
ratio, $R$) and calculated correction, $P_{EC}$,
for $0^{+} \rightarrow 0^{+}$ transitions.
The other calculated corrections, $\delta_R$ and $\delta_C$,
are given in Tables \protect\ref{radct} and \protect\ref{deltct} respectively.
\label{Exres} }
\vskip 1mm
\begin{tabular}{|rcccccc|}
\hline 
& \raisebox{0pt}[13pt][0pt]{$Q_{EC}$} &
\raisebox{0pt}[13pt][0pt]{$t_{1/2}$} &
\raisebox{0pt}[13pt][0pt]{$R$} &
\raisebox{0pt}[13pt][0pt]{$P_{EC}$} &
\raisebox{0pt}[13pt][0pt]{$ft$} &
\raisebox{0pt}[13pt][0pt]{$\F t$} \\
 & (keV) & (ms) & (\%) & (\%) & (s) & 
(s) \\[1mm]
\hline 
\raisebox{0pt}[13pt][0pt]{$^{10}$C} &
\raisebox{0pt}[13pt][0pt]{1907.77(9)} &
\raisebox{0pt}[13pt][0pt]{19290(12)} &
\raisebox{0pt}[13pt][0pt]{1.4645(19)} &
\raisebox{0pt}[13pt][0pt]{0.296} &
\raisebox{0pt}[13pt][0pt]{3038.7(45)} &
\raisebox{0pt}[13pt][0pt]{3072.9(48)} \\
$^{14}$O & 2830.51(22) & 70603(18) & 99.336(10) &
0.087 & 3038.1(18) &
3069.4(26) \\
$^{26m}$Al & 4232.42(35) & 6344.9(19) & $\geq$
99.97 & 0.083 
& 3035.8(17) &
3071.4(22) \\
$^{34}$Cl & 5491.71(22) & 1525.76(88) & $\geq$
99.988 & 0.078 
& 3048.4(19) &
3070.6(25) \\
$^{38m}$K & 6044.34(12) & 923.95(64) & $\geq$
99.998 & 0.082 
& 3049.5(21) &
3070.9(27) \\
$^{42}$Sc & 6425.58(28) & 680.72(26) & 99.9941(14)
& 0.095 
& 3045.1(14) &
3075.7(24) \\
$^{46}$V & 7050.63(69) & 422.51(11) & 99.9848(13) &
0.096 & 3044.6(18) &
3074.4(27) \\
$^{50}$Mn & 7632.39(28) & 283.25(14) & 99.942(3) &
0.100 & 3043.7(16) &
3072.9(28) \\
$^{54}$Co & 8242.56(28) & 193.270(63) & 99.9955(6)
& 0.104 
& 3045.8(11) &
3072.1(27) \\
 & & & & \multicolumn{2}{c}{Average,
$\overline{\F t}$} & 3072.2(8)~ \\
 & & & & \multicolumn{2}{c}{$\chi^2/\nu$}  &
0.6 \\[1mm]
\hline
\end{tabular}
\end{center}
\end{table}
}

\begin{figure}[t]
\centerline{   
\epsfxsize=12cm
\epsfbox{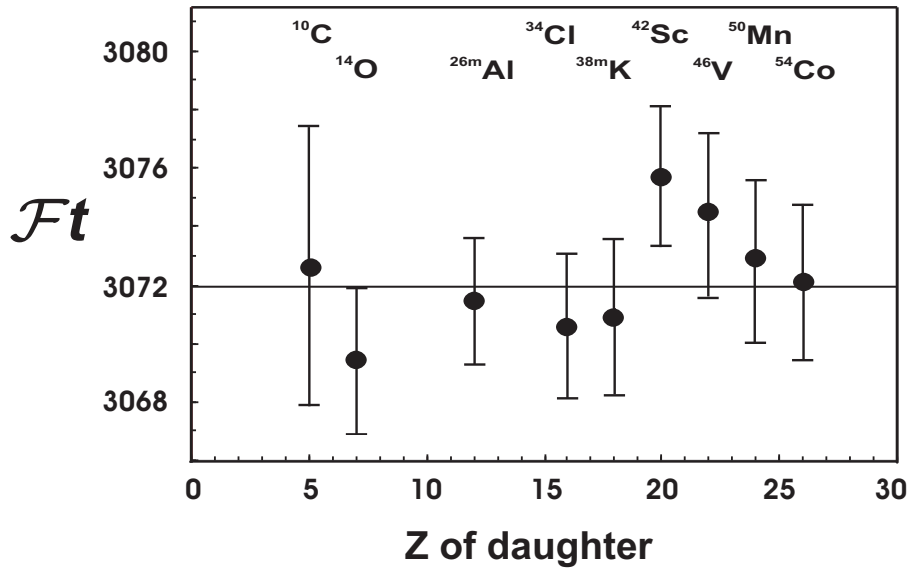}
}
\caption{$\F t$-values for the nine precision data, and the
best
least-squares one-parameter fit.
\label{fig1}}
\end{figure}

It is now convenient to separate the radiative
correction into two terms:
\be
\delta_R = \delta_R^{\prime} + \delta_{NS}
\label{deltaR}
\ee
where the first term, $\delta_R^{\prime}$, is a
function of the electron's maximum energy
and the charge of the daughter nucleus, $Z$; it therefore
depends on the particular nuclear
decay, but is {\it independent} of nuclear
structure.  To emphasize the different sensitivities of the
correction terms, we re-write
the expression for $\F t$ as
\be
\F t \equiv ft (1 + \delta_R^{\prime} )(1 + \delta_{NS} - \delta_C )
\label{Ftnew}
\ee
where the first correction term in brackets is
independent of nuclear structure, while
the second incorporates the structure-dependent terms.

To date, superallowed  $0^{+} \rightarrow 0^{+}$
transitions have been measured to $\pm 0.1\%$ precision
or better in the decays of nine nuclei ranging
from $^{10}$C to $^{54}$Co. World data on
$Q$-values, lifetimes and branching ratios -- the results
of over 100 independent measurements -- were
thoroughly surveyed \cite{ist:Ha90} in 1989 and then
updated several times since then, most recently at
the time of the last WEIN conference \cite{TH98}.  
These results appear in the first three columns of Table
\ref{Exres}.  Using the calculated electron-capture
probabilities \cite{ist:Ha90} given in the next column,
we obtain the ``uncorrected" $ft$-values listed in
column 5 with partial half-lives determined from the
formula $t = t_{1/2}(1+P_{EC})/R$.

We save a detailed discussion of the radiative and
isospin symmetry-breaking corrections for Sec.\ \ref{Sect3}.  In the
present context, though, it should be noted that the
values for $\delta_R$ and $\delta_C$ used to derive $\F t$ were taken from the last column of
Tables \ref{radct} and \ref{deltct}, respectively; each value results from more than one
independent calculation.  In
the case of $\delta_R$, the calculations are in complete
accord with one another; for $\delta_C$, we have used
an average of two independent calculations with
assigned uncertainties that reflect the (small) scatter
between them.  Thus, in a real sense, both
experimentally and theoretically, the $\F t$-values
given in  Table \ref{Exres} and plotted in  Fig.\
\ref{fig1} represent the totality of
current world knowledge.  The uncertainties reflect the
experimental uncertainties and an estimate of the {\em
relative} theoretical uncertainties in $\delta_C$  There is
no statistically significant evidence of inconsistencies in
the data ($\chi^2 / \nu =
1.1$), thus verifying the expectation of CVC at the level
of $3 \times 10^{-4}$, the fractional uncertainty quoted
on the average $\F t$-value.

In using this average $\F t$-value to determine
$\GV$, we must account for additional uncertainty:
{\em viz}

\be
\overline{\F t} = 3072.3 \pm 0.9 \pm 1.1 ,
\label{Ftavg1}
\ee

\noindent where the first error is the statistical error of
the fit (as illustrated in Fig.\ \ref{fig1}),
and the second is an error related to the systematic
difference between
the two calculations of $\delta_C$ by Towner and Hardy \cite{TH02}
and by Ormand and Brown \cite{OB95} that we have
combined in reaching
this result.  (For a more complete discussion of
how we treat these theoretical uncertainties, see
reference \cite{ist:Ha90}.)  We now add the two
errors linearly to obtain the value we use in
subsequent analysis:

\be
\overline{\F t} = 3072.3 \pm 2.0 .
\label{Ftavg2}
\ee

The value of $V_{ud}$ is obtained by relating
the
vector
constant,
$\GV$, determined from this $\overline{\F t}$
value, to
the weak
coupling constant from muon decay, $\GF /(\hbar
c)^3
=
(1.16639 \pm 0.00001) \times 10^{-5}$
GeV$^{-2}$, according to:

\be
V_{ud}^2 = \frac{K}{2 \GF^2 (1+\DRV )
\overline{\F
t}} .
\label{Vud2f}
\ee

\noindent The result obtained is

\be 
\mids V_{ud} \mids \, = 0.9740 \pm 0.0005 ,
~~~~~~~~~~~~~~~~~~~~[{\rm Nuclear}]
\label{Vud00}
\ee

\noindent where the nucleus-independent
radiative
correction has
been set at

\be
\DRV = (2.40 \pm 0.08) \% .
\label{DRV}
\ee

\noindent Note this value differs slightly (but
within
errors) from an 
earlier value \cite{MS86} because of the decision
by Sirlin \cite{Si94} to centre the cut-off parameter
$\mA$, where $(m_{a_1}/2) \leq \mA \leq 2
m_{a_1}$, exactly at the $a_1$-meson mass
when
evaluating the axial contribution to the
radiative-correction
loop graph.

From the value of $V_{ud}$ given in Eq.\
(\ref{Vud00}),
the unitarity
sum, Eq.\ (\ref{Unitarity}), becomes

\be
\sum_i V_{ui}^2 = 0.9968 \pm 0.0014 ,
~~~~~~~~~~~~~~~~~~~[{\rm Nuclear}]
\label{Unit00}
\ee

\noindent which fails to meet unity by 2.2
standard
deviations.  In
connection with this result, we note the following
two
points:

(a)  The error bar associated with $ \, \mids V_{ud}
\mids \,$ in Eq.\
(\ref{Vud00}) is {\em not}
predominantly experimental in origin.  In fact, if
experiment were the sole contributor, the
uncertainty would
be only $\pm 0.0001$.  The largest contributions
to the $ \, \mids V_{ud} \mids \,$ error bar come
from $\DRV $ ($\pm 0.0004$) and $\delta_C$
($\pm 0.0003$).

(b)  The unitarity result in Eq.\ (\ref{Unit00})
depends on the values of
nuclear structure-dependent
corrections.
In Sec.\ \ref{Sect3} we will examine whether the
failure
to meet
unitarity can be repaired by reasonable
adjustments to
these
corrections.  Our conclusion is largely negative. 
Other speculative
possibilities are presented in Sec.\ \ref{Sect4}.

\subsection{Neutron decay} \label{neutd}

On the one hand, free neutron decay has an
advantage over nuclear decays since
there are no nuclear-structure dependent
corrections to
be
calculated.
On the other hand, it has the disadvantage
that it is not purely vector-like but has a mix of
vector and axial-vector contributions.
Thus, in addition to a lifetime measurement,
a
correlation 
experiment is also required to separate the
vector and
axial-vector pieces.  Both types of
experiment present serious experimental
challenges.
The value of $V_{ud}$ is determined from the
expression

\be
V_{ud}^2 = \frac{K / \ln 2}{\GF^2 (1 + \DRV )
(1 + 3
\lambda^2 )
f (1 + \delta_R ) \tau_n } ,
\label{Vud2n}
\ee

\noindent where $\lambda$ is the ratio of
axial-vector
and
vector
effective coupling constants, $\lambda =
\GA^{\prime}
/
\GV^{\prime}$,
with $\GA^{\prime 2} = \GA^2 (1 + \DRA )$
and
$\GV^{\prime 2} = \GV^2 (1 + \DRV )$.  Here
$\DRA$
and $\DRV$ are the
nucleus-independent radiative corrections.  With
the
experimental
measurement of $\lambda$ being a determination
of
$\GA^{\prime} / \GV^{\prime}$, the actual
value of
$\DRA$ is
not  required for the evaluation of $V_{ud}^2$.
In Eq.\ (\ref{Vud2n}), $f$ is the statistical rate
function
and
$\delta_R$, the nucleus-dependent radiative
correction
evaluated for
the case of a neutron.  Wilkinson \cite{Wi82}
has evaluated the product
$f ( 1 + \delta_R )$.  His value was revised by
Towner
and Hardy \cite{TH95}
who incorporated the current best $Q$-value to
obtain
$f ( 1 + \delta_R ) = 1.71489 \pm 0.00002$.
Lastly, $\tau_n$ is the
mean
lifetime for
neutron decay.

{\footnotesize
\begin{table}[t]
\begin{center}
\caption{Experimental results$^{a}$ for neutron
decay
\label{exptneut}}
\vskip 1mm
\begin{tabular}{|ccllll|}
\hline  
& \raisebox{0pt}[13pt][0pt]{Method} &
\multicolumn{2}{c}{\raisebox{0pt}[13pt][0pt]{Measured
values }}
&~&
\multicolumn{1}{c|}{\raisebox{0pt}[13pt][0pt]{Average }} \\[1mm]
\hline  
\raisebox{0pt}[13pt][0pt]{$\tau_n$(s)} &
\raisebox{0pt}[13pt][0pt]{$n$ beam} &
\raisebox{0pt}[13pt][0pt]{$~~891 \pm 9$ \cite{Sp88}} & &
&  \\
& $n$ trap & $~~877 \pm 10$
\cite{Pa89}  
& $~~887.6 \pm 3.0$ \cite{Ma89} 
& & \\
& & $~~888.4 \pm 3.3$ \cite{Ne92} 
& $~~882.6 \pm 2.7$ \cite{Ma93} 
& & \\
& & $~~889.2 \pm 4.8$ \cite{By96}  
& $~~885.4 \pm 1.0$ \cite{Ge98} 
& & \\
& \multicolumn{2}{l}{Overall Average} & & & $~~885.6 \pm 0.8$ \\[3mm]
$\lambda$ & $\beta$-asym. & $-1.254 \pm
0.015^b$
\cite{Kr75} &
$-1.257 \pm 0.012^b$ \cite{Er79}  
& & \\
& & $-1.262 \pm 0.005$ \cite{Bo86} &
$-1.2594 \pm 0.0038$ \cite{Ye97}  & & \\
& & $-1.266 \pm 0.004$ \cite{Li97} & $-1.2739
\pm
0.0019$ \cite{Ab02}
& & \\
& $\beta$ and $\overline{\nu}$ asym & $-1.2686 \pm
0.0047$ \cite{Mo01}
& & & \\
& $\beta$-$\overline{\nu}$ corr. & $-1.259 \pm
0.017$
\cite{St78} & & & \\
& \multicolumn{2}{l}{Overall Average} & & & $-1.2690 \pm 0.0022$ \\[1mm]
\hline 
\end{tabular}
\end{center}

\noindent\footnotesize 
{$^{a}$ Following the practice used for 
 superallowed decay, we retain only those
measurements
with uncertainties
that are within a factor of ten of the most precise
measurement for
each quantity.  All such measurements, of which we are 
aware, that have not been withdrawn
by their authors are listed.}

\noindent\footnotesize {$^b$ Corrected for
weak
magnetism and recoil 
following ref.\,\cite{Wi82}.} 

\end{table}    
}

A survey of world data on neutron decay appears in
Table \ref{exptneut}.  The lifetime measurements,
when considered as a single body of data, are
statistically consistent.  However the measurements
of $\lambda$ are not ($\chi^2 / \nu = 2.6$) and, as a
consequence, the uncertainty quoted in the table for
the overall average value of $\lambda$ has been
scaled by a factor of 1.6.  Inserting the average
values for $\tau_n$ and $\lambda$
into Eq.\ 
(\ref{Vud2n}), we determine the value of
$V_{ud}$ to be

\be 
\mids V_{ud} \mids \, = 0.9745 \pm 0.0016 ,
~~~~~~~~~~~~~~~~~~~~[{\rm Neutron}]
\label{Vudnavg}
\ee

\noindent and the unitarity sum to be

\be
\sum_i V_{ui}^2 = 0.9978 \pm 0.0033 ,
~~~~~~~~~~~~~~~~~~[{\rm Neutron}]
\label{Unitnavg}
\ee

\noindent a value that agrees with unitarity {\em
and} with the nuclear result, Eq.\ 
(\ref{Unit00}) which is over a factor of two more
precise.  In connection with this result, we note
the following two points:

(a) For neutron decay, the error bar associated with $
\, \mids V_{ud} \mids \,$ in Eq.\
(\ref{Vudnavg}) is some three times larger than
the error bar obtained from nuclear decays, Eq.\
(\ref{Vud00}); however, in contrast with the
latter case, it is predominantly experimental in
origin.  The largest theoretical contribution to
the $ \, \mids V_{ud} \mids \,$ error bar comes (at
the level of $\pm 0.0004$) from $\DRV $, a
correction that is common to both nuclear and
neutron decays.

(b) Currently, the theoretical uncertainty on
$\DRV $ dominates the nuclear result for
$\, \mids V_{ud} \mids \,$.  As experimental
results for the neutron improve, $\DRV $ will
eventually dominate the neutron result too. 
Therefore, so long as $\DRV $ remains at its
current level of uncertainty, the neutron results
will never be able to test unitarity with
significantly better precision than the nuclear
decays do now, in spite of their independence of
$\delta_C$.  They will, of course, be able to
provide an important test of whether there are
some systematic problems with the
nuclear-dependent corrections, which are
not now anticipated
in the theoretical uncertainty quoted in Eq.\
(\ref{Vud00}).

\subsection{Pion beta decay} \label{pibd}

Like neutron decay, pion beta decay has an
advantage
over nuclear
decays in that there are no nuclear
structure-dependent
corrections
to be made.  It also has the same advantage as the nuclear
decays in
being a
purely
vector transition, in its case $0^{-} \rightarrow 0^{-}$, so
no separation of vector and
axial-vector components
is required.  Its major disadvantage,
however, is that 
pion beta decay, $\pi^{+} \rightarrow \pi^0
e^{+}
\nu_e$,
is a very weak branch, of the order of $10^{-8}$.  This
results in severe experimental limitations.  For the pion
beta decay, the value of $V_{ud}$ is determined from the
expression

\be
V_{ud}^2 = \frac{K / \ln 2}{\GF^2 (1 + \DRV )
f_1
f_2 f
(1 + \delta_R)
\tau_m / BR },
\label{Vud2pi}
\ee

\noindent where $f$ is the approximate
statistical
rate
function
\be
f = \frac{1}{30} \left ( \frac{\Delta}{m_e}
\right
)^5
\label{pif}
\ee
\noindent with $\Delta$ being the pion mass
difference 
($\Delta = m_{\pi^{+}} - m_{\pi^{0}}$) and
$m_e$, the
electron mass.
The mass difference is known with high precision
from
the
work
of Crawford \etal \cite{Cr91} to be
$\Delta = ( 4.5936 \pm 0.0005)$ MeV.
The factors $f_1$ and $f_2$ are corrections to
$f$, and
are
easily calculable functions \cite{Mc85} of
$\Delta /
m_{\pi^{+}}$
with values of $f_1 = 0.941039$ and $f_2 =
0.951439$.
The nucleus-dependent radiative correction
evaluated
for
the
case of the pion is $\delta_R = (1.05 \pm 0.15 )
\% $.
Finally, $\tau_m$ and $BR$ are the pion mean
lifetime
and
branching ratio, respectively.

Two precise lifetime
measurements \cite{Ko95,Nu95}
were published in 1995 and the PDG02 average
is

\be
\tau_m = (2.6033 \pm 0.0005 ) \times 10^{-8}
~{\rm s}
.
\label{pitau}
\ee

\noindent The branching ratio is principally from
McFarlane
\etal \cite{Mc85}:

\be
BR = (1.025 \pm 0.034) \times 10^{-8} .
\label{piBR}
\ee

\noindent Inserting Eqs.\ (\ref{pitau}) and
(\ref{piBR})
into
Eq.\ (\ref{Vud2pi}), we determine the value of
$V_{ud}$ 
to be

\be 
\mids V_{ud} \mids \, = 0.9670 \pm 0.0161 ,
~~~~~~~~~~~~~~~~~~~~[{\rm Pion}]
\label{Vudpi}
\ee

\noindent and the unitarity sum

\be
\sum_i V_{ui}^2 = 0.9833 \pm 0.0311 ,
~~~~~~~~~~~~~~~~~~[{\rm Pion}]
\label{Unitpi}
\ee

\noindent satisfying the unitarity condition but with
comparatively large uncertainty.  The
error
on
$\, \mids V_{ud} \mids \,$
is entirely due to the uncertainty in the pion
branching
ratio.
There is an experiment currently
underway at PSI designed to
improve
this
branching ratio by a factor of eight.  If
successful, the
error
on $\, \mids V_{ud} \mids \,$ would be reduced to $
\pm
0.0022$, still less precise than the current
results from both neutron and nuclear decays.  Of course,
ultimately it
too will be limited by the theoretical uncertainty on
$\DRV$. 

\section{Theoretical corrections in nuclear
decays}
\label{Sect3}

In Sec.\ \ref{nsFd} we noted that the value of
$V_{ud}$
determined
from nuclear decays -- the most precise result
available --
resulted in the unitarity test
among 
the elements of the first row of the CKM matrix
not
being 
satisfied by two standard deviations.  Here, we
discuss
the
theoretical corrections involved in the determination,
and assess
whether the
failure to meet unitarity can be
removed by reasonable adjustments in
these
calculations.    
To restore unitarity, the calculated radiative
corrections for all nine nuclear transitions
would have to be shifted downwards by 0.3\%
($\ie$ as
much as one-quarter of their current value), or
the
calculated
isospin symmetry-breaking correction shifted upwards by 0.3\%
(over 
one-half their value), or some combination of
the
two. 

\subsection{Radiative corrections}
\label{rc}

As mentioned in Sec.\ \ref{nsFd}, the radiative
correction is
conveniently
divided into
terms that are nucleus-dependent, $\delta_R$,
and terms
that are not,
$\DRV$.  These are written

\bea
\delta_R & = & \delta_R^{\prime} + \delta_{NS}
\nonumber \\
\delta_R^{\prime} & = & \frac{\alpha}{2 \pi} \left [
\overline{g}(E_m)
+ \delta_2 + \delta_3  \right ]
\nonumber \\
\DRV & = & \frac{\alpha}{2 \pi} \left [ 4 \ln
(\mZ
/m_p )
+ \ln (m_p / \mA )
+ 2 C_{\rm Born} \right ] + \cdots ,
\label{drDR}
\eea

\noindent where the ellipses represent further
small
terms
of order
0.1\%.  In these equations, $E_m$ is the
maximum
electron energy
in beta decay, $\mZ$ the $Z$-boson mass,
$\mA$ the
$a_1$-meson mass, and
$\delta_2$ and $\delta_3$ the order-$Z
\alpha^2$
and
-$Z^2 \alpha^3$
contributions respectively.  The function
$g(E_e,E_m)$, which is a function of electron
energy, was first defined by Sirlin \cite{Si67} as
part of the order-$\alpha$ universal photonic
contribution arising from the weak vector current;
it is here
averaged
over
the
electron spectrum to give $\overline{g}(E_m)$. 
Finally, the terms $C_{\rm Born}$ and
$\delta_{NS}$ come from the
order-$\alpha$ axial-vector photonic
contributions: the former accounts
for single-nucleon contributions, while the latter
covers two-nucleon contributions and is
consequently dependent on nuclear structure.

\begin{table} [t]
\begin{center}
\caption{Calculated nucleus-dependent radiative
correction,
$\delta_R$, in percent units, and the component
contributions
as identified in Eq.\ (\protect\ref{drDR}).
\label{radct} }
\vskip 1mm
\begin{tabular}{|rccccc|}
\hline 
& \raisebox{0pt}[13pt][0pt]{
 $\frac{\alpha}{2\pi } \overline{g} (E_m)$}
& \raisebox{0pt}[13pt][0pt]{
 $\frac{\alpha}{2\pi } \delta_2$}
& \raisebox{0pt}[13pt][0pt]{
 $\frac{\alpha}{2\pi } \delta_3$}
& \raisebox{0pt}[13pt][0pt]{
 $\delta_R^{\prime}$}
& \raisebox{0pt}[13pt][0pt]{
 $\delta_{NS}$} \\[1mm]
\hline 
\raisebox{0pt}[13pt][0pt]{$^{10}$C} &
\raisebox{0pt}[13pt][0pt]{1.468} &
\raisebox{0pt}[13pt][0pt]{0.182} &
\raisebox{0pt}[13pt][0pt]{0.005} &
\raisebox{0pt}[13pt][0pt]{1.65(1)} &
\raisebox{0pt}[13pt][0pt]{$-$0.360(35)} \\
$^{14}$O & 1.286 & 0.227 & 0.008 &
 1.52(1) & $-$0.250(50) \\
$^{26m}$Al & 1.110 & 0.325 & 0.021 &
 1.46(2) & ~~0.009(20) \\
$^{34}$Cl & 1.002 & 0.388 & 0.034 &
1.42(3) & $-$0.085(15) \\
$^{38m}$K & 0.964 & 0.413 & 0.041 &
1.42(4) & $-$0.100(15) \\
$^{42}$Sc & 0.939 & 0.448 & 0.049 &
1.44(5) & ~~0.030(20) \\
$^{46}$V & 0.903 & 0.468 & 0.057 &
1.43(6) & $-$0.040(7)~ \\
$^{50}$Mn & 0.873 & 0.494 & 0.065 &
1.43(7) & $-$0.042(7)~ \\
$^{54}$Co & 0.843 & 0.507 & 0.073 &
1.42(7) & $-$0.029(7) \\[1mm]
\hline
\end{tabular}
\end{center}
\end{table}

Calculated values for all four components of
$\delta_R$ are given in Table \ref{radct}.  There
have been two independent calculations \cite{Si87,JR87}
of both $\delta_2$ and
$\delta_3$; they are completely consistent with
one another if proper account is taken of
finite-size effects in the nuclear charge
distribution.  The
values listed in Table \ref{radct} are our
recalculations \cite{ist:Ha90} using the formulas
of Sirlin \cite{Si87} but incorporating a Fermi
charge-density distribution for the nucleus.  Note
that we have followed Sirlin in assigning an
uncertainty equal to $(\alpha /2\pi ) \delta_3$ as an
estimate of the error made in stopping the
calculation at that order.  Also
appearing in the table are our adopted values
for $\delta_{NS}$, which we have taken from our
recent re-evaluation \cite{TH02}.

In assessing the changes in $\delta_R$ that would be
required in order to restore unitarity, it is helpful
to rewrite Eq.\ 
(\ref{drDR}) in terms of the typical values taken
by its components:  {\em viz} 

\be
\delta_R \simeq 1.00 + 0.40 + 0.05 .
\label{tdr}
\ee

\noindent Thus, $\delta_R^{\prime} \simeq 1.4\%$.
If the failure to obtain
unitarity
in the
 CKM matrix with
$V_{ud}$ from nuclear beta decay is due to the
value
of
$\delta_R^{\prime}$,
then {\it all} $\delta_R^{\prime}$ values must
be reduced to $\sim 1.1$\%.  This
is not
likely.
The leading term, 1.00\%, involves standard
QED and
is
well
verified.  The order-$Z\alpha^2$ term, 0.40\%,
while
less
secure
has been calculated twice \cite{Si87,JR87}
independently, with
results in accord.

Taking a similar approach for the
nucleus-independent radiative
correction, we write

\be
\DRV = 2.12 - 0.03 + 0.20 + 0.1\%
~~\simeq~~
2.4\% ,
\label{tdrv}
\ee

\noindent of which the first term, the
leading
logarithm,
is
unambiguous.
Again, to achieve unitarity of the
CKM matrix,
$\DRV$
would have to be 
reduced to 2.1\%: \ie\, all terms other
than the
leading
logarithm must
sum to zero.  This also seems unlikely.

\subsection{Isospin symmetry-breaking corrections}
\label{c}

Because the leading terms in the radiative
corrections
are so
well founded, attention has focussed more on
the
isospin symmetry-breaking
correction.  Although smaller than the radiative
correction,
the isospin symmetry-breaking correction is clearly sensitive to
nuclear-structure issues.  It comes about because
Coulomb
and charge-dependent nuclear forces destroy
isospin
symmetry between the initial and final states in
superallowed beta-decay.  The consequences are
twofold:
there are different degrees of configuration
mixing in
the
two states, and, because their binding energies
are not
identical, their radial wave functions differ. 
Thus,
we
accommodate both effects by writing $\delta_C
=
\delta_{C1} + \delta_{C2}$.  Constraints can be
placed
on
the calculation of $\delta_{C1}$ by insisting
that
it reproduce the measured coefficients of the
isobaric
mass
multiplet equation.  Constraints are placed on
$\delta_{C2}$
by insisting that the asymptotic forms of the
proton and
neutron
radial functions match the known separation
energies.

\begin{table} [t]
\begin{center}
\caption{Calculated isospin symmetry-breaking correction,
$\delta_C$, in
percent units.
\label{deltct} }
\vskip 1mm
\begin{tabular}{|rccccc|}
\hline 
& \raisebox{0pt}[13pt][0pt]{TH$^a$}
& \raisebox{0pt}[13pt][0pt]{OB$^b$}
& \raisebox{0pt}[13pt][0pt]{SVS$^c$}
& \raisebox{0pt}[13pt][0pt]{NBO$^d$}
& \raisebox{0pt}[13pt][0pt]{Adopted$^e$} \\
 & & & & & Value \\[1mm] 
\hline 
\raisebox{0pt}[13pt][0pt]{$^{10}$C} &
\raisebox{0pt}[13pt][0pt]{0.18} &
\raisebox{0pt}[13pt][0pt]{0.15} &
\raisebox{0pt}[13pt][0pt]{0.00} &
\raisebox{0pt}[13pt][0pt]{0.12} &
\raisebox{0pt}[13pt][0pt]{0.17(3)} \\
$^{14}$O & 0.32 & 0.15 & 0.29 &      & 0.24(3) \\
$^{26m}$Al & 0.27 & 0.30 & 0.27 &      &
0.29(3)
\\
$^{34}$Cl & 0.64 & 0.57 & 0.33 &      & 0.61(3)
\\
$^{38m}$K & 0.62 & 0.59 & 0.33 &      & 0.61(3)
\\
$^{42}$Sc & 0.49 & 0.42 & 0.44 &      & 0.46(3)
\\
$^{46}$V & 0.43 & 0.38 &      &      & 0.41(3) \\
$^{50}$Mn & 0.51 & 0.35 &      &      & 0.43(3) \\
$^{54}$Co & 0.61 & 0.44 & 0.49 &      & 0.53(3)
\\[1mm]
\hline 
\end{tabular}
\end{center}

\noindent\footnotesize {~~~~~~~~~~~~~~~~~~$^a$
Ref. \cite{TH02}
;~~~~$^b$
Ref. \cite{OB95}
;~~~~$^c$
Ref. \cite{SVS96} ;
~~~~$^d$ Ref. \cite{NBO97} .}

\noindent\footnotesize {~~~~~~~~~~~~~~~~~~$^e$
Average of OB
and
TH; assigned uncertainties reflect the}

\noindent\footnotesize
{~~~~~~~~~~~~~~~~~~~~{\em relative}
scatter between these calculations.}
\end{table}

The results of several calculations for
$\delta_C$ are shown in Table \ref{deltct}.  The
values in the first column are those calculated
by the methods developed by Towner, Hardy
and Harvey \cite{THH77} and refined in more
recent publications \cite{To89,Ha94,TH02}.  They
result from shell-model calculations to
determine $\delta_{C1}$, and full-parentage
expansions in terms of Woods-Saxon radial wave
functions to obtain $\delta_{C2}$.  Ormand
and Brown, whose values \cite{OB95} for
$\delta_C$ appear in column 2, also employed
the shell model for calculating $\delta_{C1}$
but derived $\delta_{C2}$ from a self-consistent
Hartree-Fock calculation.  Both of these
independent calculations for $\delta_C$
reproduce the measured coefficients of the
relevant isobaric multiplet mass equation, the
known proton and neutron separation energies,
and the measured $ft$-values of the
weak non-analogue $0^{+} \rightarrow
0^{+}$ transitions \cite{Ha94} where they are known.  In
our analysis in Sec.\ \ref{nsFd}, we have
used the average of these two
sets of $\delta_C$ values: our adopted
values
appear in the last column of the table.

Two more recent calculations provide a valuable
check that these $\delta_C$ values are not
suffering from severe systematic effects. 
Sagawa, van Giai and Suzuki \cite{SVS96}
have added RPA correlations to a
Hartree-Fock
calculation that incorporates charge-symmetry
and charge-independence breaking forces in the
mean-field potential to take account of isospin
impurity in the core; the correlations, in
essence,
introduce a
coupling
to the isovector monopole giant resonance.  The
calculation is not constrained, however, to
reproduce known
separation energies.
Finally, a large shell-model calculation has been
mounted
for the
$A=10$ case by 
Navr\'{a}til, Barrett and Ormand \cite{NBO97}. 
Both
of
these two
new works have 
produced values of $\delta_C$ very similar to,
but actually
{\em smaller} than those used in our analysis,
\ie\
worsening
rather
than helping the unitarity problem.

The typical value of $\delta_C$ is of order
0.4\%. 
If the
unitarity
problem is to be solved by improvements in
$\delta_C$,
then {\it all}
$\delta_C$ values must be increased so that
the typical value is raised to around 0.7\%. 
There is
no
evidence
whatsoever for such a shift from recent works.

\section{Speculative suggestions to resolve unitarity
problem}
\label{Sect4}

Since it is concluded in Sec.\ \ref{Sect3} that
reasonable
adjustments
to the theoretical corrections, $\delta_R$,
$\DRV$ and $\delta_C$, are unlikely to
resolve the
unitarity
problem posed by the nuclear result for
$V_{ud}$,
we turn in this section to some more speculative
alternatives.  We discuss four suggestions, two that
do not require extensions to the Standard Model and
two that do.  All but one require the introduction of
important new physics to explain the apparent
discrepancy.

\subsection{Saito-Thomas correction}
\label{STc}

One suggestion to resolve the
unitarity
problem was
made some years ago by Thomas \cite{ST95}.  It is based on a
quark-meson
coupling model,
in which nuclear matter consists of
non-overlapping
nucleon (MIT)
bags bound by the self-consistent exchange of
$\sigma$
and $\omega$
mesons in the mean-field approximation.  The
model is
extended
to include an isovector-vector meson ($\rho$)
and an
isovector-scalar meson ($\delta$).  The coupled
equations
to be
solved are 
very similar to those of Quantum
Hadrodynamics \cite{SW86}, but
involve boundary conditions at the bag radius. 
As a
consequence
of a coupling to meson fields the quark mass in
a
medium
becomes
an effective one,

\be
m_i^{\ast} = m_i - ( V_{\sigma} \pm
\sfrac{1}{2}
V_{\delta} ),
~~~~~~~~~~i = u,d
\label{effqm}
\ee

\noindent with the upper sign for the up quark. 
Here
$V_{\sigma}$
and $V_{\delta}$ are strengths of
$\sigma$-meson and
$\delta$-meson
mean fields.  At the quark level, the conserved
vector
current (CVC)
hypothesis is broken if the up and down quarks
have
different masses,
but for a free nucleon this breaking is second
order in
$(m_d - m_u)$.
In a nuclear medium, however, this breaking
may
become
first
order in $(m_d^{\ast} - m_u^{\ast})$, but a
critical
constraint
of the calculation is to show it reverts to $(m_d -
m_u)^2$ at
zero density.

Saito-Thomas (ST) solve the coupled equations
in an
infinite nuclear-matter approximation, the
solutions being
functions of
the
matter 
density, $\rho_B$.  Let $\psi_{i/j}$ be the
wavefunction
of quark $i$
bound in nucleon $j$ in nuclear matter; then, the
following overlap
integral can be defined between quark $i$ bound
in a
proton and
quark $i^{\prime}$ bound in a neutron:

\be
I_{i i^{\prime}}(\rho_B ) = \int_{{\rm Bag}}
dV
\psi_{i/p}^{\dag}
\psi_{i^{\prime}/n} .
\label{qoverlp}
\ee

\noindent Calculations indicate that $I_{i
i^{\prime}}(\rho_B
)$ varies
linearly in $\rho_B$, being unity at zero density. 
What
is
required
for beta decay is the product of three overlap
integrals

\be
\mids I_{ud} \mids^2 \times 
\mids I_{uu} \mids^2 \times 
\mids I_{dd} \mids^2 \equiv 1 - \delta_C^{{\rm
quark}}
\label{dcquark}
\ee

\noindent and this product also is approximately
linear
in
the density.
The results of the calculations \cite{ST95} are

\be
\delta_C^{{\rm quark}} = b \times \left (
\frac{\rho_B}{\rho_0} \right ) ,
\label{dcquark1}
\ee

\noindent where $\rho_0$ is the saturation
density for
equilibrium
nuclear matter, and $b$ is in the range $(0.15 -
0.20)\%$
for bag
radii in the range $(0.6 - 1.0)$ fm.  Thus
$\delta_C^{{\rm quark}}$,
at densities of $\rho_0/2$, lies in the range
0.075\% to
0.10\%
while, at $\rho_0$, it lies between 0.15\% and
0.20\%.  If such a correction were to be applied
to the results from nuclear superallowed transitions, 
the corresponding
value for $V_{ud}^2$ would then be
increased
by
these amounts.

However, in their original work, the
authors themselves admit \cite{ST95} that their
results are merely qualitative, having been
derived from a model that deals only with
nuclear matter.  More recently, they have also
recognized \cite{Th98} that a different model actually produces
a considerably smaller effect.  Furthermore, there remains the
question of whether a simple addition of the
Saito-Thomas
correction,
$\delta_C^{{\rm quark}}$,
to the $\delta_C$ already computed in a
nucleons-only
calculation is the correct approach at all, since it
may lead
to double counting.  In the nucleons-only case,
certain
parameters
are adjusted to reproduce Coulomb observables
such as
the $b$- and 
$c$-coefficients of the isobaric mass multiplet
equation,
and proton
and neutron separation energies.  Thus, the
parameters
become effective ones, 
which in some unquantifiable way, actually
contain quark
effects.

In summary, the Saito-Thomas correction
continues to be unspecified for finite nuclei and
now appears for several reasons
to be a much smaller effect than was originally proposed.
It seems clear that this effect, if it is applicable
at all, would be insufficient in itself to
restore unitarity.  

\subsection{Ad hoc parameterized fit}
\label{Twopf}

Wilkinson \cite{Wi90} has suggested that 
data like those
displayed in
Fig.\ \ref{fig1} might be better fitted by a
quadratic function
in $Z$,
representing a
further correction of speculative origin.  He considers the value
of this fitted function at $Z = 0$ to be the
appropriate $\F t$-value to take forward in the determination
of $V_{ud}$.  Alternatively, he has also considered
the option of disregarding the structure-dependent
corrections entirely and then fitting the uncorrected
$ft$-values with a similar function quadratic in $Z$.  There is no
physical justification whatsoever for incorporating
any Z-dependent corrections beyond those already
accounted for in $\delta_R$ and $\delta_C$. 
Furthermore, there is no statistically significant
indication from the $\F t$-value data in Fig.\ \ref{fig1}
that one is required.  Consequently, until such
time as new $\F t$-value data can demonstrate a clear residual
$Z$-dependence, this suggested explanation
for the
nuclear unitarity result must be regarded as
unsatisfactory.

\subsection{Right-hand currents}
\label{RhandC}

Because experimental evidence at low energies
favors maximal parity violation in weak interactions,
that condition has been built into the Standard
Model.  However, one class of possible extensions to
the Standard Model would restore parity symmetry
at higher energies through the introduction of
additional heavy charged gauge bosons that are
predominantly right-handed in character.  Under
certain specific conditions, such models could
explain the nuclear results described in Sec.\
\ref{nsFd}.

For example, an extension known as the manifest
left-right symmetric model \cite{BBMS77} leads to
a revised form \cite{TH95} for Eq.\ (\ref{Vud2f})
which includes a mixing angle $\zeta$:

\be
V_{ud}^2 (1-2\zeta) = \frac{K}{2 \GF^2 (1+\DRV )
\overline{\F
t}} .
\label{RHCVud2f}
\ee

\noindent If, under these conditions, we require that
$V_{ud}^2$ must satisfy unitarity, {\em viz}

\be
V_{ud}^2 = 1-V_{us}^2-V_{ub}^2,
\label{RHCUn}
\ee

\noindent then, with the $\overline{\F t}$ value
taken from Eq.\ (\ref{Ftavg2}), we can derive a
value for the mixing
angle of $\zeta = 0.0015 \pm
0.0007$.  Within the context of the manifest
left-right symmetric model, this is a fully
satisfactory result; not surprisingly, it is two
standard
deviations away from the Standard Model's
pure $V-A$ value.

\subsection{Scalar interaction}
\label{Scalint}

If the Standard Model of pure $V-A$ weak
interactions is
extended instead to include
scalar and tensor interactions, then the
expression for
the beta
spectrum shape would include an additional term
coming from
the scalar
interaction, which is inversely proportional to the
electron
energy.  Since the $\F t$ values include an integral
over the spectrum shape, the presence of a non-zero
scalar interaction would be reflected by a correction
to the $\F t$ values that is inversely proportional to
the decay energy of the parent nucleus.  To test this
possibility, we fit the data in Fig.\ \ref{fig1} by the
function

\be
\F t = k \left [ 1 + b_F \gamma \langle W^{-1}
\rangle
\right ] ,
\label{FtbF}
\ee

\noindent where $k$ and $b_F$ are parameters
determined from the fit.  The latter, $b_F$, is
known as the Fierz interference term; a non-zero
value for this term would signal the presence of a
scalar interaction. In Eq.\ (\ref{FtbF}), $\gamma^2
= ( 1 - (\alpha Z)^2)$,
$\alpha$ is
the
fine-structure
constant, $Z$ the atomic number of the daughter
nucleus,
and
$\langle W^{-1} \rangle$ is the value of $1/W$
averaged
over the
electron spectrum, where $W$ is the electron
energy in
electron
rest-mass units.  The value obtained for the Fierz
interference term from the data in Fig.\ \ref{fig1} is

\be
b_F = -0.0027 \pm 0.0029 .
\label{bFvalu}
\ee

\noindent A negative sign is expected \cite{JTW57} 
for a positron emitter.
The 90\% confidence level upper limit is

\be
\mids b_F \mids \, < 0.0075 .
\label{bFupper}
\ee

\noindent To proceed to $V_{ud}$, we need a
value of
$\F t$ extrapolated to the $Z = 0$ limit.  To this
end, we fit the nine values of $\gamma \langle
W^{-1} \rangle$ by a second-order polynomial in
$Z$ and use this polynomial to provide the value of
$\gamma \langle W^{-1} \rangle$ at $Z = 0$.  With
$k$ and $b_F$ taken from the fit, 
Eq.\
(\ref{bFvalu}), we obtain

\bea
\F t(0) & = & 3067.5 \pm 2.9
\nonumber \\
\mids V_{ud} \mids \, & = & 0.9748 \pm
0.0006
~~~~~~~~~~~~~~~~~~~~[{\rm
Nuclear~+~bF}]
\nonumber \\
\sum_i V_{ui}^2 & = & 0.9983 \pm 0.0016 .
~~~~~~~~~~~~~~~~~~~[{\rm Nuclear~+~bF}]
\label{VudbF}
\eea

\noindent This result is now in accord with
unitarity but, of course, requires the presence of a
non-zero scalar coupling to achieve that goal.

We close this section by noting that all four
suggested explanations we have presented for the
non-unitarity result in Eq.\ (\ref{Unit00})
are
entirely speculative
and should be considered with caution.

\section{Future experimental prospects}
\label{tfw}

It is evident from the foregoing discussion that the
current world data on $V_{ud}$ are tantalizingly
close to producing a definitive result on unitarity. 
The nuclear measurements have achieved the
highest experimental precision but they are now
constrained by theoretical uncertainties.  The
neutron and pion measurements are, as yet,
experimentally less precise than the nuclear ones,
but they are free from one of the more important
sources of theoretical uncertainty, $\delta_C$.  All
three classes of measurements are now being
extended and improved at a number of laboratories,
and there are good prospects for considerably
reduced error bars within a few years.  In
combination with the re-visitation of the $K_{e3}$
decay \cite{Sh02}, these results should settle the
uncertainty over $\delta_C$ and determine whether
the deviation from unitarity, apparent from the
nuclear result, is real or not.  This is ample
argument to justify considerable experimental
activity.  At the same time, it should not be
forgotten that the full impact of these experimental
advances will be diluted until there are theoretical
improvements in the correction $\DRV$ -- the
dominant source of theoretical uncertainty in all
cases.  Ultimately, to make significant
improvements in the unitarity test, there will have to
be advances in both theory and experiment.

Of the three classes of experiment, that focusing on
pion decay is currently farthest from the precision
required for a meaningful unitarity test.  We noted
in Sec.\ \ref{pibd} that a considerable improvement
is anticipated in the foreseeable future, but the result
will not reach a precision higher than
that already achieved for neutron decay.  There is
considerable optimism, however, that the neutron
measurements themselves can be improved as new
experiments with ultra-cold neutrons come to
fruition. The outcome can be expected seriously to
challenge the nuclear experiments for experimental
precision, but will take at least a few more years to
do so.  

As to the nuclear experiments, the nine
superallowed transitions whose $ft$
values are
known to within a fraction of a percent
have been the subject of intense scrutiny
for at least the past three decades.  All except
$^{10}$C have the special advantage that the
superallowed branch from each is by far the
dominant transition in its decay ($>$ 99\%). 
This
means that the branching ratio for the 
superallowed transition can be
determined to high precision from relatively
imprecise measurements of the other weak
transitions, which can simply be
subtracted from 100\%.  Given the quantity of
careful measurements already published, are
there reasonable prospects for significant
improvements in these decay measurements in
the near
future?  Given the uncertainty in the
theoretical corrections, which experiments can
shed the most light on the efficacy of these
corrections?

If we begin by accepting that it is valuable
for experiment to be at least a factor of two
more precise than theory, then an examination
of the world data shows that the Q-values for
$^{10}$C, $^{14}$O, $^{26m}$Al and
$^{46}$V, the half-lives of $^{10}$C,
$^{34}$Cl and $^{38m}$K, and the
branching ratio for $^{10}$C can all bear
improvement.  Such improvements will soon
be feasible.  The Q-values will reach 
the required level (and more) as mass
measurements with new on-line Penning traps
become possible; half-lives will likely yield to
measurements with higher statistics as 
high-intensity beams of separated isotopes are
developed for the new radioactive-beam
facilities; and, finally, an improved 
branching-ratio measurement on $^{10}$C has
already
been made with Gammasphere and simply
awaits analysis \cite{FR98}.

Qualitative improvements will also come
as we increase the number of superallowed
emitters 
accessible to precision studies.  The greatest
attention recently has been paid to the $T_z =
0$ emitters with $A \geq 62$, since these
nuclei are expected to be produced at new
radioactive-beam facilities, and their
calculated structure-dependent corrections,
$\delta_C$ - $\delta_{NS}$, (see Eq.(\ref{Ftnew}))
are predicted to be large \cite{OB95,SVS96,TH02}.
They could then
provide a valuable test of the accuracy of
$\delta_C$ calculations.  It is likely, though,
that the required precision will not be
attainable for some time to come.  First, with
current Penning-trap technology, these nuclei are too short
lived ($t_{1/2} \leq 100$ms) for their masses to be
measured with the required precision.  Second, the decays
of these nuclei are of higher energy and
each must involve numerous weak Gamow-Teller
transitions that together will account for
significant intensity in addition
to the superallowed transition \cite{HT02}.  Branching-ratio 
measurements will thus be very
demanding, particularly with the limited
intensities likely to be available initially for
these rather exotic nuclei.  Finally, lifetime
measurements will often be similarly constrained by
statistics, although not in every case.  The half-life
of $^{74}$Rb was recently obtained \cite{Ba01} with fully
adequate precision.

More accessible in the short term will be
the $T_z = -1$ superallowed emitters with $18
\leq A \leq 38$.  There is good reason to
explore them.  For example, the calculated
value \cite{TH02} of the total structure-dependent
correction, $\delta_C$ - $\delta_{NS}$, for
$^{30}$S
decay,
though smaller than the values expected
for the heavier nuclei, is actually 1.13\% -- 
about a factor of two larger than for any other
case currently known -- while $^{22}$Mg
has a rather low value of 0.51\%.  If such large
differences are confirmed by the measured
$ft$-values, then it will do much to increase
our confidence in the calculated isospin symmetry-breaking
corrections.   To be sure, these decays will
provide a challenge, particularly in the
measurement of their branching ratios, but the
required precision should be achievable with
isotope-separated beams that are currently
available.  In fact, such experiments are already
underway at the Texas A\&M cyclotron.  Results for
the decay of $^{22}$Mg are currently being prepared
for publication.

\section{Conclusions}
\label{conc}

The current world data on superallowed $0^+
\rightarrow 0^+$ beta decays lead to a
self-consistent
set of $\F t$-values that agree with CVC but
differ
provocatively, though not yet definitively, from
the
expectation of CKM unitarity.  There are no
evident
defects in the calculated radiative and isospin symmetry-breaking
corrections that could remove the problem, so, if
any
progress is to be made in firmly establishing (or
eliminating) the discrepancy with unitarity,
additional
experiments are required.  We have indicated
what
some relevant nuclear experiments might be.

In the past decade there have been significant
improvements in
the measurements of the neutron lifetime and
beta
asymmetry,
and further improvements are promised in the
near
future.
It is likely that these studies in neutron decay
will
soon
approach the results from nuclear superallowed
decays
in precision; and when there, they will have the
advantage
in that their results are not dependent on
nuclear-structure
corrections.  So far, though, the neutron result
is quite consistent with the nuclear one.

Clearly, there is strong motivation to pursue
experiments, not only on
the neutron and nuclear front but also in re-visiting
the $K_{e3}$ decay,
since, if firmly established, a discrepancy with
unitarity
would
indicate the need for important new physics.

\section*{Acknowledgments}

We would like to thank Bill Marciano for drawing our
attention to new experimental results on the $K_{e3}$ decay.
The work of JCH was supported by
the U.S. Department of Energy under Grant
number DE-FG03-93ER40773 and by the Robert
A. Welch Foundation; he would also like to thank the
Institute for Nuclear Theory at the University of Washington
for its hospitality and support during part of this work.

\end{document}